# When, why and how to test spreadsheets

Louise Pryor
115 Hanover Street, Edinburgh EH2 1DJ, UK
www.louisepryor.com

**ABSTRACT**

*Testing is a vital part of software development, and spreadsheets are like any other software in this respect. This paper discusses the testing of spreadsheets in the light of one practitioner's experience. It considers the concept of software testing and how it differs from reviewing, and describes when it might take place. Different types of testing are described, and some techniques for performing them presented. Some of the commonly encountered problems are discussed.*

## 1  INTRODUCTION

Any software engineer will tell you that testing is a vital part of software development. It's usually the only way to tell that a software system actually does what it is meant to. In this paper I discuss my experience of testing spreadsheets, the techniques I have used and the problems I have encountered. Many developers of spreadsheets have had little or no exposure to software engineering concepts, and may not be aware of some of the simple techniques that can be used. The emphasis is on the practical aspects of testing spreadsheets rather than the theory, and there is no intention of providing a comprehensive review. My experience is primarily based on financial modelling spreadsheets, and the techniques I use may not be appropriate in other situations.

In the remainder of this section I introduce the notion of testing, compare it to code reviewing and describe when it might take place. The remaining sections cover the different types of testing, and I conclude with a section on common problems. Throughout the paper I discuss only the testing of spreadsheet formulae, omitting all consideration of macros and tools such as Excel's solver and goal seek. Many of the techniques I describe are demonstrated in a sample Excel workbook available for download (Pryor 2004a). Some of them use XLSior, an Excel add-in I have developed to support good practice in spreadsheet development (XLSior).

### 1.1  What is testing?

Testing is the controlled execution of a spreadsheet, checking that what it does meets the specification. Typically this means using known inputs and checking the results against what would be expected from these inputs.

Testing is the only way to tell what the spreadsheet actually does. It is a vital step in the process of gaining confidence in the results (Pryor, 2004b). With a good testing process, tests are easy to run (so are run often), and logs are generated automatically, so that it is easy for a user to see what tests have been run and when, and what the results were.

### 1.2  Review is not testing

Testing is part of ensuring spreadsheet quality and complements spreadsheet review. The distinction between the two is an important one. A review consists of looking at the code and trying to spot the errors, while to test a program you run it and look at the results. Both are necessary, and neither is likely to find all the errors on its own.

A truly effective spreadsheet review would follow the lines of the systematic code inspections originally devised by Fagan and described by Gilb (1988). Although effective, this is time consuming and requires more than one person. Moreover, spreadsheet code is notoriously difficult to review, including as it does

- Spreadsheet formulae and layout
- Data validation
- Conditional and other formatting
- Defined names
- Charts
- VBA
- Definitions for the solver, scenarios, data filtering, pivot tables and other tools

Automated tools can provide valuable help, but although their use is often referred to as "testing" they are limited to summarising information and detecting common sources of error, such as formulae not being copied correctly or numbers treated as text. These types of error are essentially syntactic. A spreadsheet that passes all the tests of an automated tool may still be incorrect if its semantics do not agree with its specification.

However thorough the reviewer, in the end they cannot do more than say that in their opinion running the spreadsheet will result in the correct behaviour. The end result of a review depends on the reviewer's understanding of both the specification and, crucially, the implementation. Testing, on the other hand, may check only certain execution paths, but a successful test of those paths is much more objective than a review. Devising the tests requires some understanding of the specification, but a detailed understanding of the program logic is not necessary. The only way to tell whether a spreadsheet is producing the correct behaviour to run it; in other words to test it. Of course, the successful running of a few tests does not guarantee that the spreadsheet will produce the correct results for all possible inputs.

**1.3 When to test**

Testing can take place at different stages of the development process. The different types of testing are described in the standard software engineering books such as (Pressman & Ince 2000).

- **Unit testing** is the most detailed type of testing. Individual components are tested in isolation. Unit testing should take place frequently throughout the development process.

- **System testing** looks at the system as a whole, testing the final results. It should take place at a minimum when a spreadsheet is released for use, and preferably more often during development.

- **Regression testing** compares the results of a new version against those of a previous version. It is a specialised form of system testing, used to check that no unintended changes have been introduced.

- **Acceptance testing** is testing by the spreadsheet user (or on their behalf) when they receive the spreadsheet from the developer, to determine that it meets their requirements and is fit for use.

As usual with spreadsheets, these distinctions are less clear cut than with traditional software. The user is often the same as the developer, so there may well be no separate stage of acceptance testing. Because of the lack of modularity in spreadsheet design unit testing consists of testing individual calculations or tables of calculations. For a small spreadsheet, unit testing and system testing are essentially the same process. Acceptance testing may be performed by independent third parties, such as consultants who are asked to review a spreadsheet and pronounce on its fitness for purpose.

## 2  UNIT TESTING

The idea of unit testing is to make sure that individual calculations are correct. Unit testing is especially valuable when it can be performed automatically and often, as it can then be used to check for unanticipated side effects of changes (Astels, 2003).

### 2.1 Testing for invariants

A widely used form of unit testing is the cross checking of column and row totals; other similar tests include checking that a column of percentages add to 100%, or that specific values are always positive (or negative). These are tests for invariants; conditions that should always be true. They are run every time the spreadsheet is calculated, so the results are always up to date. It is also easy to summarise the results of these tests in one area of the spreadsheet, so that they can all be checked at once.

|   | W | X | Y | Z |
|---|---|---|---|---|
|   |   | $f_x$ =IF(X65=X64, "Pass", "Fail") | | |
|   | Static tests | | | |
|   | Sum of total cashflow = sum of individual cashflows | | | |
|   | Actual | -150.00 | 3631.86 | |
|   | Expected | -150.00 | 3631.86 | |
|   | Result | Pass | Pass | |

### 2.2 Using data tables

Most testing, though, requires that the spreadsheet be run with specially chosen input values, and the results checked against those that are expected. A number of useful tests can be performed by changing just one or two values on the spreadsheet. For example, if a discount rate is set to zero then the present value of a series of payments is the same as the sum of the amounts paid.

Data tables (in Excel) or multiple operations (in Open Office) can be used for these tests. They can be used to test several different calculations that depend on the same value, as in the example below. Depending on the calculation settings they are kept up to date as the spreadsheet is recalculated. The cell whose changing value is investigated by the table may contain a complex formula, allowing the isolation of a single calculation step.

| Wage infl | Constr wages | Op wages | max factor | min factor |
|---|---|---|---|---|
| | -410.52 | -4074.23 | 11.742721 | 1.002466 |
| 0% | -400.00 | -1749.00 | 1 | 1 |
| Expected | -400.00 | -1749.00 | 1 | 1 |
| Result | Pass | Pass | Pass | Pass |

If wage inflation is zero, total construction wage costs are -1 * months * base rate. Similarly for operation wage costs. Also max wage inflation factor should be 1; so should min.

Formula bar: {=TABLE(,O15)}

The use of data tables in Excel has some disadvantages. First, the cell containing the input value must be on the same sheet as the data table. This means that you cannot have a single sheet (or set of sheets) that contain all your tests, but must scatter the tests throughout the workbook. Also, the presence of many data tables may slow down the spreadsheet calculation. In Excel you can set the calculation to Automatic (except tables), but this means that the tests are not automatically kept up to date, and there is no visible indication that this is the case.

| Wage infl | Constr wages | Op wages | max factor | min factor |
|---|---|---|---|---|
| | -369.47 | -4074.23 | 11.742721 | 1.002466 |
| 0% | -400.00 | -1749.00 | 1 | 1 |
| Expected | -360.00 | -1749.00 | 1 | 1 |
| Result | **Fail** | Pass | Pass | Pass |

**2.3 Macros**

A data table can be used either to test the effect of a single value on several results, or to test the effect of a pair of values on a single result. It cannot be used when the test depends on a large set of test data, such as setting whole ranges to zero, one, or other specific values.

A more flexible solution is to use a macro that will substitute in the appropriate values, recalculate the spreadsheet and record the results. The substitutions can be performed on individual cells or on whole ranges. For example, many of the spreadsheets I encounter perform their calculations in columns. The cells in a single column contain the same (relative) formula. That formula can be tested by substituting appropriate values in the columns forming the inputs to the formula. Often, setting the inputs to zero or one provides useful information. It is possible to test the validity of the calculations in each column using this technique.

I use the functionality provided by XLSior, which allows you to specify data substitutions that should be made and conditions that should then be true. It records the substitutions and conditions (and the results) for each test, providing an auditable record of the testing that has been performed.

| Status: | Passed |
| --- | --- |
| Run at: | 2004-04-10 16:29 |

| stitute n | Notes | Data range | Value type | Value | |
| --- | --- | --- | --- | --- | --- |
| 0 | | Inflationpa | Direct | 0 | |
| 0 | | WageInflation | Direct | 0 | |

| ie2 | Outcome | Condition | Result of Value1 | Result of Value2 | Result of condition |
| --- | --- | --- | --- | --- | --- |
| -360 | Passed | SUM('Project Cashflows'!D31:D1030)=-ConstructionWages*ConstructionMonths | -360 | -360 | TRUE |
| -1749 | Passed | SUM('Project Cashflows'!E31:E1030)=- | -1749 | -1749 | TRUE |

XLSior also provides a summary of all the tests with their results. The tests do not update automatically whenever the workbook is recalculated, but must be specifically run. Because the records include a time stamp, it is easy to tell whether the test results are up to date.

| | Tests last run at: | 2004-04-10 16:29 | | |
| --- | --- | --- | --- | --- |
| | Number of tests run: | 5 | | |
| | Number passed: | 4 | | |
| | Number failed: | 1 | | |

| Id | Description | Status | Sheet | Cell |
| --- | --- | --- | --- | --- |
| 1 | Zero wage inflation | Passed | X~TestProject | $A$12 |
| 2 | dividends of 1 per month | Passed | X~TestProject | $A$27 |
| 3 | Test construction months | Passed | X~TestProject | $A$40 |
| 4 | Lookups | Failed | X~TestMisc | $A$12 |
| 5 | Sumproduct | Passed | X~TestMisc | $A$33 |

A system of macros such as this allows flexible unit testing that is easy to reproduce. Whenever a change is made, it is easy to rerun all the tests and make sure that there are no unintended side effects.

**2.4 Devising unit tests**

The efficacy of testing depends on the tests that are used. When devising tests, you should try to break the spreadsheet. What might cause it to go wrong? In devising unit tests you are trying to isolate specific calculations or parts of calculations. For instance, suppose you have a stream of cash flows at irregular intervals. Tests might include making the intervals regular, and making all the cash flows the same amount. You should always try to test boundary values, for example setting interest rates to zero.

Other useful tests include making sure that the correct row (or column) is being found when using a lookup function. You can do this by setting the contents of the lookup table to known values, such as 11 in the first row and column, 23 in the second row and third column, and so on. Similarly, setting all elements except one of a range to zero can help in making sure that sumproducts are aligned correctly.

More generally, you should try to include tests that exercise every branch of a conditional, such as complex if statements or formulae involving maxima or minima.

## 3 SYSTEM TESTING

System testing involves testing the whole spreadsheets. Typically, you test a range of scenarios, covering both typical inputs and unusual combinations. To do system testing properly, you need an independent calculation of the expected results. Such independent calculations are, in my experience, rarely available and so full system testing is not often performed.

The only effective way of running reliable system tests is to use macros. The process is very similar to the regression testing described in the next section. The chief difficulty is determining what the correct answers should be.

## 4 REGRESSION TESTING

Regression testing is a special kind of system testing, in which the independent calculations are performed by a different version of the same spreadsheet (or sometimes by a precursor system). The chief aim of regression testing is to check that the results have not been altered by changes to the spreadsheet, or, where they have altered, to investigate the effects of the changes. It is important when adding new functionality to a spreadsheet, to make sure that the existing functionality is not affected. Regression testing on its own is not sufficient to test the correctness of a spreadsheet, as the older version against which the comparison is being made might itself contain errors.

Regression testing is really only feasible when the layout of the outputs does not change between versions, or when there are only a few outputs that need to be compared. It is more difficult to compare whole sets of calculations because of the difficulty of expressing the correspondences between the cells in the two versions. It is particularly useful for spreadsheets that are used to calculate results in a standard format, possibly for use in another program.

A typical set of regression tests would consist of a number of scenarios that cover a wide range of the possible inputs. The old and new versions are both run on each scenario and the results compared. It is often possible to reduce the time that the testing takes by running the old version through each scenario once and recording the results, although this is only necessary if the calculations are very slow.

I have usually found it necessary to write custom macros for each spreadsheet that is to be tested. The testing is usually performed in a third spreadsheet. It is made significantly easier if the important results on the spreadsheet to be tested are on a single sheet. I set up a testing spreadsheet with four sheets for each scenario: one to hold the input values, two to hold the results (one from each version), and one to hold the comparison between the results.

The macro I use usually has the following form
```
Open the old and new workbooks
For each scenario
    Copy the inputs to both versions
    Recalculate both versions
    Copy the results to the results sheets
Next scenario
```

I usually use the Import feature in XLSior, which records the source and time of the imports, to copy the results. The comparison sheet can be used to check that the absolute values of the results are the same, or that they are within a specified tolerance.

## 5    PROBLEMS

Although systematic, automated testing is a useful tool in dealing with spreadsheets, the ride is not always smooth. I use unit testing techniques a lot, and regression testing when appropriate, but rarely do full independent system tests. Even with unit tests, there are a number of problems that are encountered regularly.

- It is often difficult to devise effective tests if there is no specification other than the spreadsheet itself. This is not a problem when developing a spreadsheet from scratch, but can be a serious obstacle if you are using tests to help review someone else's spreadsheet. The tests soon become tautologous if the formulae are being used to derive the specification that is used to drive the tests for the formulae.

- Dealing with floating point arithmetic is sometimes tricky. It is often necessary to use a pair of inequalities instead of a single equality.

- If a test fails, it may be a problem with the test rather than with that which is being tested. This can be very frustrating.

- It can be very time consuming to set up a full range of tests. I usually address this by not attempting to do so at the outset. I start with a few tests of significant calculations, and gradually add more tests. Typically I add new tests for any changes that I make. Whenever an error is found I add a test that would have detected it, and other tests that would detect similar errors. And sometimes I just add more tests when I have time, or as an idea occurs to me for a good one.

- It is often difficult to devise tests if there are large, complex formulae. The use of simpler formulae with explicit intermediate results makes testing much easier, as you can test the derivation of the intermediate results independently of each other and of the way in which they are combined.

None of these problems is specific to spreadsheets, and I encounter them regularly when testing systems written in more conventional programming languages.

## CONCLUSIONS

Systematic testing of spreadsheets is both possible and desirable. Of the problems that are encountered, some (such as the difficulty of regression testing when the layout has changed) are peculiar to spreadsheets while others (the care needed when dealing with floating point arithmetic, lack of specification) occur whatever the programming language used. There are some simple techniques that can be used to set up automated tests, not all of which require macros.